\documentclass{aastex}
\usepackage{spr-astr-addons}
\usepackage{url}

\RequirePackage{color}

\begin{document}

\title{Modelling the thermal absorption and radio spectra of the binary pulsar B1259$-$63}
\slugcomment{Not to appear in Nonlearned J., 45.}
%% Running heads
\shorttitle{Short article title}
\shortauthors{Autors et al.}

\author{O. Koralewska\altaffilmark{1}} 
\and 
\author{J. Kijak\altaffilmark{1}}
\and 
\author{W. Lewandowski\altaffilmark{1}}
\altaffiltext{1}{Janusz Gil Institute of Astronomy, University of Zielona G\'ora,
 ul.  Szafrana 2, 65-516 Zielona G\'ora, Poland
 }

\begin{abstract}
We present the results of modelling of the radio spectrum evolution and dispersion measure variations of PSR~B1259$-$63, a pulsar in a binary system with Be star LS~2883. 

We base our model on a hypothesis that the observed variations of the spectrum are caused by thermal free-free absorption occurring in the pulsar surroundings. We reproduce the observed pulsar spectral shapes in order to examine the influence of the stellar wind of LS~2883 and the equatorial disc on the pulsar's radiation.

The simulations of the pulsar's radio emission and its consequent free-free absorption give us an insight into the impact of stellar wind and equatorial disc of LS~2883 has on the shapes of PSR~B1259$-$63 radio spectra, providing an evidence for the connection between gigahertz-peaked spectra phenomenon and the close environment of the pulsar. Additionally, we  supplement our model with an external absorbing medium, which results in a good agreement between simulated and observational data. 
\end{abstract}


%\section*{}
%\label{sec:intro}

\section{Introduction}
The pulsar B1259$-$63, which was discovered in 1989 by \cite{johnston92a}, is a first known radio pulsar in a binary system with a massive, main-sequence star. Radio timing observations of this 48-ms pulsar and optical observations of its companion, a Be star LS~2883, allowed for the determination of the orbital parameters of the binary system. The orbital period of the system is 1237~days, and the orbit of the neutron star is highly eccentric ($e \approx 0.87$) with a projected semi-major axis $ a \sin i \approx 1296 \textrm{ light s}$ and a longitude of periastron $ \omega \approx 138.65^{\circ}$. The inclination of the orbit was estimated from mass function using the pulsar mass of $M_{\textrm{p}}=1.4 \textrm{M}_{\odot}$ and the Be star mass $M_{\textrm{Be}}=10 \textrm{M}_{\odot}$ and resulted in the inclination angle $ i = 36^{\circ}$ \citep{johnston94}. One of the most distinguishable features of the system is the fact that PSR~B1259$-$63 is eclipsed for about $\sim 35$~days when being close to the periastron, which is explained by the presence of a dense circumstellar disc around LS~2883. This disc is inclined with respect to the orbital plane \citep{melatos95}. The last detection of PSR~B1259$-$63, before it crosses the disc, occurs roughly 18~days prior to the periastron passage, then the pulsed emission reappears on the 15th~day (at the frequency 3.1~GHz) or on the 16th~day (at the frequency 1.4~GHz) after the periastron passage \citep{chernyakova2014}. Besides creating the circumstellar disc, the strong outflow of particles from LS~2883 is a source of a hot, polar wind, that also interacts with pulsar radiation, causing its absorption, but also directly collides with the pulsar wind, generating shock-powered X-ray emission \citep{hirayama99}. 

The observed emission from this interesting system covers a wide range of the electromagnetic spectrum, from radio waves to high-energy gamma-rays, showing strong variability as the pulsar orbits its companion \citep{chernyakova2014}. In our studies, we concentrated on the PSR~B1259$-$63 radio spectrum and its shape changes that appear to be related to the orbital phase. A typical pulsar radio spectrum can be described by a power-law function with average value of spectral index $<\alpha>~\approx{-1.6}$ (\citealt{lorimer95}, \citealt{maron2000}, \citealt{jankowski2017}). Several pulsars exhibit high frequency turnovers in their spectra at much higher frequencies around $\sim$ 1~GHz. These objects have been called gigahertz-peaked spectra (GPS) pulsars. Currently, there are 17 known objects of this type \citep{kijak2017}. It was suggested that their spectral shape is caused by the influence of a pulsar's environment on its radiation \citep{kijak2011b}. The case of PSR~B1259$-$63 is found to be essential for understanding the origins of GPS behaviour. The analysis of its spectrum evolution reveals similarities with high frequency turnover phenomenon. The main hypothesis that attempts to explain
the observed non-typical spectral shapes is the thermal free-free absorption \citep{kijak2011a, lewandowski2015} and even our simulations based on the simplified model of the stellar wind alone \citep{dembska2015} seem to support this idea, and thus indicate that the GPS-type spectra can be explained using the same phenomenon for both, binary and isolated pulsars. Studies of the relation between the observed pulsar spectrum shape and the pulsar orbital phase may provide useful information on the gigahertz peaked spectra of pulsars. For most GPS pulsars their environmental conditions are more stable than in the case of PSR~B1259$-$63 and their spectra shape doesn't change with time. The significance of phase-dependent variations in PSR~B1259$-$63 spectra lies in the ability to link the shape of high-frequency turnovers with the changes of pulsar surroundings, resulting from various path lengths to the observer. An analysis of that relation helps to explain the observed non-typical spectral shapes at least qualitatively, as it was done by \citet{dembska2015}. Furthermore, the comparison between simulated and observational data gives us an opportunity to estimate physical parameters of absorbing medium that is affecting the shape of the PSR~B1259$-$63 radio spectrum.
 
In this paper we present the results of our simulations of the variability of the PSR~B1259$-$63 spectrum. We have modeled the geometry of the binary system based on the data available for this system. The model we developed here is more adequate than the previous version presented in \citet{dembska2015}, which included the thermal absorption in the spherically symmetrical stellar wind, neglecting the presence of an equatorial disc. The simplified model was able to recreate the basic properties of the PSR~B1259$-$63 spectrum evolution, i.e., an increase of the amount of the absorption with decreasing distance from periastron, while the maximum flux shifts towards higher frequencies. However, in contrast to the observational spectra, the simulation results obtained by \citet{dembska2015} showed almost no absorption for the orbital phases far from the periastron passage. Furthermore, the simulated spectra, unlike the observed ones, were symmetrical with respect to the periastron point, which was caused by not accounting for the observed value of the longitude angle $\omega = 138.65 ^\circ$ \citep{johnston94}. The simplified model of the thermal absorption in the LS~2883 stellar wind allows us to estimate the optical depth of the absorbing medium. This physical quantity depends on both electron density and temperature. Without the independent estimation of one of these parameters their influences on the PSR~B1259$-$63 spectra are inseparable.

Here we combined both the observations of the spectrum of the pulsar at various orbital phases, as well as the dispersion measure variations $\Delta DM$ reported in the literature. Moreover, in our model we considered the influence of both the polar stellar wind and the equatorial disc of LS~2883. This is - to our knowledge - the most detailed model that was used to explain the variations of both the $DM$ and the observed flux densities of PSR~B1259$-$63.

%We showed, that the high-frequency turnovers in spectra of B1259$-$63 for orbital phases close to pulsar's eclipse are caused by the influence of the circumstellar disc. The loss of energy in spectra corresponding to pulsar's greater distances from periastron couldn't be recreated by modelling the influence of the stellar wind alone while maintaining the agreement between observational and simulated values of $DM$ variations. 

%-------------------------------------------------------------------

\section{A model of PSR~B1259$-$63/LS~2883 system}

The radio spectrum of PSR~B1259$-$63 exhibits variations that were shown to coincide with the orbital phase change. During pulsar's orbital motion, depending on the actual orbital phase, its radiation travels trough stellar wind and/or the broad circumstellar disc, and both the geometrical travel distance, as well as the optical depth along the line-of-sight will vary. Furthermore, the absorbing medium interacting with pulsar's radiation is characterized by different physical parameters (namely the electron temperature and density). These parameters will vary both along the line-of sight - since 
it will cross regions with different properties - as well as in time, since the line-of-sight itself will be moving according to the pulsar motion.

We have investigated the influence of thermal free-free absorption on the pulsar's spectrum shape by solving the equation of radiative transfer (e.g. \citealt{rohlfs2004}):
\begin{equation}
I_{\nu}(s)=I_{\nu}(0)e^{-\tau_{\nu}(s)}+B_{\nu}(T)(1-e^{-\tau_{\nu}(s)}),
\label{radiative_transfer}
\end{equation}
where $B_{\nu}(T)$ is a Planck function, here equal to Rayleigh-Jeans approximation $2({\nu}^2/{c^2})kT$, $I_{\nu}(0)$ is substituted with the intrinsic flux $S_{\nu}(0)$ and $I_{\nu}(s)$ with the total measured flux $S_{\nu}(s)$. Since the intrinsic flux $S_{\nu}(0)$ is inaccessible to observations, for purposes of our simulations, we assumed that $S_{\nu}(0)$ of PSR~B1259$-$63 can be described by a power-law function of the observation frequency $\nu$ \citep{sieber73}:
\begin{equation}
S_{\nu}(0)=A\cdot \nu^\alpha,
\label{power-law}
\end{equation}
with the spectral index value $\alpha=-1.3$ \citep{dembska2015} obtained from averaging the observational data only for frequencies 4.8~GHz and 8.6~GHz. We used only the highest frequencies to estimate the intrinsic pulsar spectrum, as these should not be affected by external effects in any significant way. Thermal absorption effect for the selected frequencies is negligible and spectral ageing of the electrons doesn't affect the flux density measurements at the frequencies below $\sim 20$~GHz \citep{chhetri2012}. In order to estimate the spectrum scaling $A$ for each spectrum separately, we fitted the simulated spectra to observations of flux densities using non-linear least squares method. 

In our model, we assumed that PSR~B1259$-$63 spectrum evolution results from the variations of the optical depth along the line-of-sight. For thermal free-free absorption the optical depth $\tau_{\nu}$ is given by \citep{rohlfs2004}:
\begin{align}
\tau_{\nu}= 3.014\cdot 10^{-2} \left( \frac{T_{e}}{\mathrm{K}}\right)^{\frac{-3}{2}} \left(\frac{\nu}{\mathrm{GHz}}\right)^{-2} \left(\frac{EM}{\mathrm{pc \cdot cm^{-6}}}\right) \cdot \nonumber \\ \cdot \left[ \mathrm{ln} \left( 4.955 \cdot 10^{-2} \frac{\nu} {\mathrm{GHz}}\right)+1.5 \mathrm{ln} \left( \frac{T_{e}}{\mathrm{K}}\right)\right]
\label{optical_depth}
\end{align}
where $T_{e}$ is the electron temperature of the absorbing medium and $EM$, and the emission measure is defined as \citep{rohlfs2004}:
\begin{equation}
\frac{EM}{\mathrm{pc}\cdot \mathrm {cm^{-6}}}=\int_{0}^{\frac{s}{\mathrm{pc}}}\left(\frac{N_{e}}{\mathrm{cm^{-3}}}\right)^{2}\mathrm{d}\left(\frac{s}{\mathrm{pc}}\right),
\label{EM}
\end{equation}
where $N_{e}$ is the electron density. 

To obtain the total optical depth along a given line-of-sight, we integrated using steps small enough for the values of $N_e$ and $T_e$ to change linearly between the steps boundaries, which allowed us to use a simple trapezoidal rule. The adopted filling factor value is equal 1, since we integrate only along the line of sight occupied by the electrons, from the pulsar's location to the 100~AU from the LS~2883 (we are interested only in DM variations, not in total DM and the low electron density of the stellar wind at the distances greater than 100~AU doesn't affect the PSR~B1259$-$63 radio spectra shape). In our model the electron density $N_e$ both in the stellar wind and disc decreases continuously with the distance from the LS~2883, we didn't consider any clumpiness in the absorbing medium.

Since the observed optical depth will be a function of both the electron density and temperature distributions within the system, it is impossible to decouple the effects of both properties in the model of the pulsar spectrum (i.e. the optical depth) only. Therefore, we decided to use additional information that was available for the PSR~B1259$-$63, namely, the observations of the dispersion measure variations. This quantity is temperature independent, at least when adopting the so-called cold-dispersion approximation. By definition $DM$ is an integrated column-density of the electrons in the intervening space between the  pulsar and observer \citep{rohlfs2004}:
\begin{equation}
\frac{DM}{\mathrm{pc}\cdot \mathrm {cm^{-3}}}=\int_{0}^{\frac{s}{\mathrm{pc}}}\left(\frac{N_{e}}{\mathrm{cm^{-3}}}\right)\mathrm{d}\left(\frac{s}{\mathrm{pc}}\right),
\label{DM}
\end{equation}

   \begin{figure}[t]
   \centering
   \includegraphics[width=0.7\hsize]{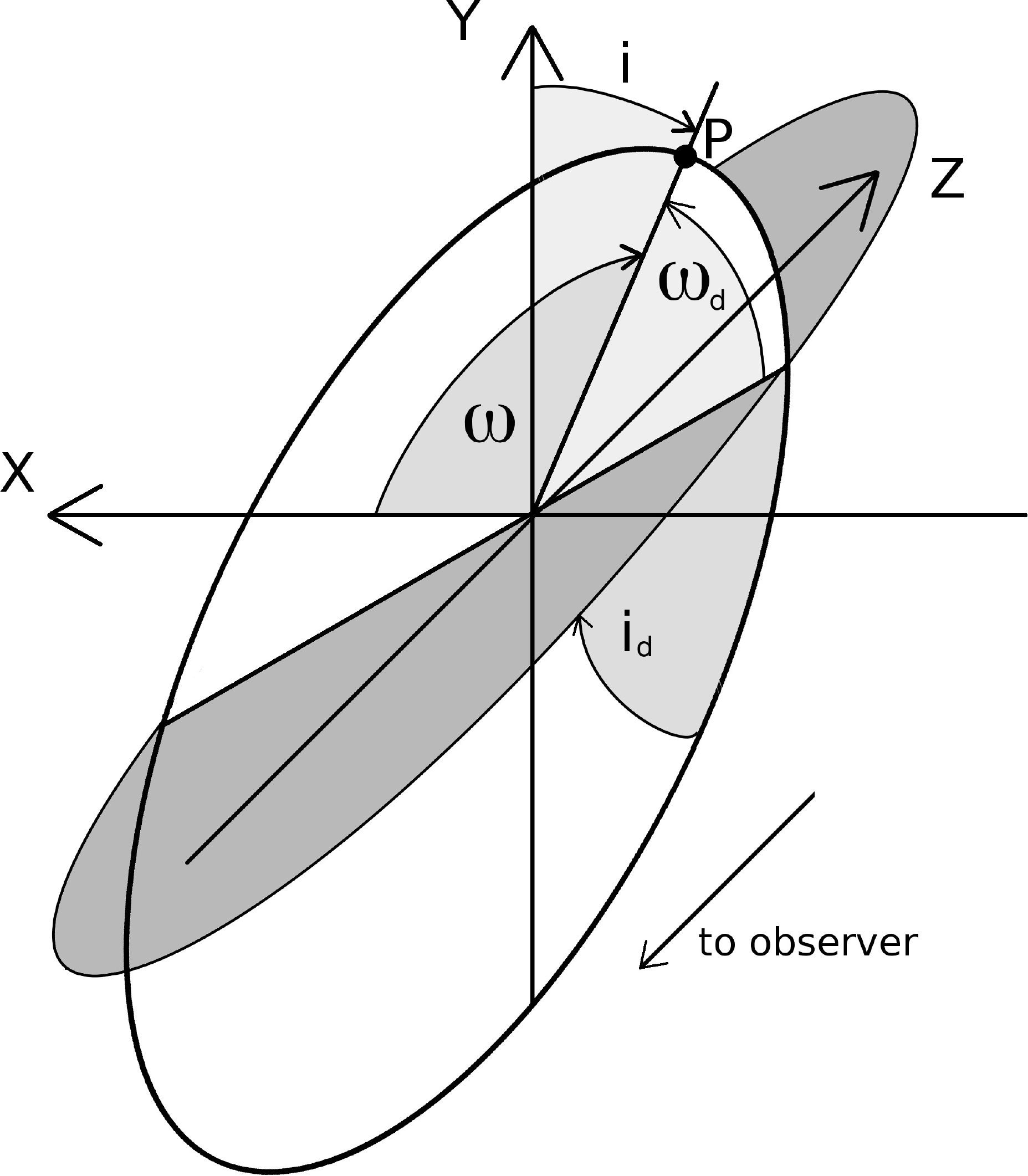}
   \caption{Schematic model of a PSR~B1259$-$63 in a binary system with LS~2883. The Be star is in the center of the coordinates system. The observer's line of sight lies along Z axis. P indicates the pulsar's periastron, $i$ is the orbital inclination with respect to the plane of the sky XY, $\omega$ is the longitude of periastron, $i_d$ is the angle between the orbital plane and the plane of the Be disc and $\omega_{d}$ is the angle between the direction of periastron and the line of intersection of the Be disc with the orbital plane.}
    \label{FigModelGeometry}
    \end{figure}

The orbital parameters of the binary system B1259\linebreak$-$63/LS~2883, the geometry of the equatorial disc and the stellar wind, as well as their physical properties, i.e., electron temperature and density distribution, create a large set of free parameters in our model (the geometry of the model is presented in Fig.~\ref{FigModelGeometry}). We reduced their number by fixing the values of parameters known from radio timing of the PSR~B1259$-$63, its eclipse and the optical observations of LS~2883 (specific values are described in the following paragraphs). Observations of dispersion measure variations \citep{wang2004} allowed us to impose restrictions on the values the electron density of both the wind and disc component ($N_{e W}(1 \textrm{AU})$ and $N_{e D}(0)$, respectively). 
The simulation results where compared to PSR~B1259$-$63 flux density measurements obtained during three orbital cycles (\cite{johnston99}, \cite{connors2002}, \cite{johnston2005}).

%-------------------------------------------------------------------

\subsection{PSR 1259-63: Orbital parameters}

In order to calculate the amount of free-free absorption of the pulsar's radiation, we need to know its path through the absorbing medium. The position of PSR~B1259$-$63 orbit, its shape and size, determine the thickness of layers interacting with radio waves emitted by pulsar. The observed and derived parameters, obtained by \citet{johnston94} that are used in our simulation, are: the projected semi-major axis of the orbit $a\sin{i} = 1295.98 \textrm{~light s}$, eccentricity of the orbit $e = 0.87$, longitude of periastron $\omega=138.65^{\circ}$ and B1259$-$63 orbital period $P_{b}=1237 \textrm{~days}$. Assuming the pulsar mass $M_{\textrm{p}}=1.4 ~\textrm{M}_{\odot}$ and Be star mass $M_{\textrm{Be}}=10 ~\textrm{M}_{\odot}$ the estimated inclination angle equals $ i = 36^{\circ}$ \citep{johnston94}. Taking this inclination into account, the length of semi-major axis of the pulsar orbit is  $a=4.42 ~\textrm{AU}$.

%-------------------------------------------------------------------

\subsection{LS~2883: Stellar wind and equatorial disc}
The stellar wind model used in our computations is spherically symmetric with a constant velocity, same as applied in the version presented
in \citet{dembska2015}. For this model, the electron density variations in the wind are a simple consequence of the conservation of mass and the inverse square law:
\begin{equation}
N_{e W}(r) = {N_{e W}(1 \textrm{AU})\over r^2},
\label{Ne_wind}
\end{equation}
where $N_{e W}(1 \textrm{AU})$ is a free parameter representing the stellar wind electron density at the distance $r=1 ~\textrm{AU}$ from the Be star. Another free parameter connected with thermal absorption in the stellar wind is the wind's electron temperature $T_{e W}$, which in our simulations is uniform. 

For the circumstellar disc, we modelled the electron density distribution using the equation
from \citet{hummel2000}:
\begin{equation}
N_{e D}(R,Z)=N_{e D}(0)R^{-m}\exp\left[-{1\over2}\left(Z\over H(R)\right)^2\right].
\label{Ne_disc}
\end{equation}
Here $R$ denotes the distance from the Be star surface $r$ relative to Be star radius $R_{\star}$: $R=r/R_{\star}$ and $Z$ is the distance to the plane of circumstellar disc $z$ given in the units of stellar radius $R_{\star}$: $Z=z/R_{\star}$. $H(R)$, the scale height of the disc, is given by \citep{hummel2000}:
\begin{equation}
H(R) = {\sqrt{{5 \over 3} R_{g} T_{d}} \over \sqrt{G M_{\star} / R_{\star} }}R^{3 \over 2},
\label{Disc_height}
\end{equation}
where the enumerator is associated with the speed of sound in mono-atomic gases, the denominator is the Keplerian orbital velocity at $R = R_{\star}$, $R_{g}$ is the universal gas constant, $M_{\star}$ is the Be star mass and $T_{d}$ is the disc temperature, which in our simulations was one of the three free parameters describing equatorial disc physical properties. The other two were disc's electron density on the surface of Be star $N_{e D}(0)$ and the index $m$ associated with the radial electron density distribution. For Be star discs the values of $m$ obtained from infrared observations are typically in the range between $2$ and $3.5$ \citep{waters86}.

The simulated circumstellar disc was inclined with respect to the plane of the orbit at the angle $i_{D}$ and also rotated by the longitude angle $\omega_{D}$ (the angle between the direction of periastron and the line of intersection of the Be disc with the orbital plane), that gave us two more free parameters specifying the geometry of the Be star disc. The disc is believed to be highly tilted with respect to the pulsar orbital plane, according to PSR~B1259$-$63 timing observations, it is also suggested that its intersection with the plane of the orbit is roughly perpendicular to the orbit's major axis \citep{johnston99}. The values of the angle between the disc's and orbit's plane $i_{D}$ giving good agreement with dispersion measure and rotation measure variations were found to be in the range from $10^{\circ}$ to $40^{\circ}$ \citep{melatos95}.
%-------------------------------------------------------------------

\subsection{Modelling procedure}
The pulsar's orbit, Be star disc and the stellar wind shapes were generated in the coordinates system associated with the observer. The first step of modelling was to position these elements of the investigated binary system by rotating them by the given angles (the geometric model is presented in the Fig.~\ref{FigModelGeometry}). Then, using the rules of the celestial mechanics, we expressed the pulsar's orbital phases in the form of the number of days prior to or after the periastron passage (denoted with minus and plus respectively, this was done to make our model compatible with the observational data). The values of the optical depth and the extra dispersion measure were initially calculated only for the lines of sight corresponding to the pulsar's orbital phases, for which we had observational data. Each of these was divided into small integration steps, and the position of the central point was used to estimate the electron density of the stellar wind and the disc separately, according to their density distribution (Eq. \ref{Ne_wind} and Eq. \ref{Ne_disc}). For the purposes of the $DM$ contribution, one only needs the total electron density, i.e., a sum of densities from both model components. The contribution of a given step to the total $EM$ value was calculated separately, since, to take it into account for the purposes of calculating the optical depth for different model components, we also had to use different electron temperatures.

Next we have integrated obtained electron densities along the line-of-sight to the observer, in order to get the values of both the emission measures and the dispersion measure $DM$ corresponding to the given orbital phases. The $EM$ value is associated with the optical depth $\tau_{\nu}$, together with electron temperature $T_{e}$ and the frequency of observation $\nu$ (Eq. \ref{optical_depth}). Setting these three values allowed us to calculate the amount of thermal absorption of the pulsar's intrinsic flux $S_{\nu}(0)$ (Eq.~\ref{power-law}) for a given orbital phase.

\section{Simulation results and discussion}

In this section we present the results obtained in our modelling $DM$ variations that allowed us to estimate electron densities of the stellar wind and disc, and also to specify the equatorial disc shape and its orientation. Since the amount of extra $DM$ caused by the free electrons within the binary system does not depend on the electron temperature of the wind (but it depends on the disc temperature), this approach allowed us to estimate the characteristic electron densities of both the stellar wind and the disc components, as well as the disc geometry and its temperature. This basically leaves only the electron temperature of the wind as a free parameter when attempting to model the observed pulsar spectra.

%-------------------------------------------------------------------

\subsection{Dispersion measure variations}

In order to estimate electron density of the stellar wind and disc, we have simulated the $DM$ variations due to the orbital phases for the binary system PSR~B1259$-$63/LS~2883. The optimal electron densities and geometric parameters were chosen based on a comparison of the simulated data with the values of  $\Delta DM$ observed during the 1997 periastron passage \citep{wang2004}. Although the dispersion measure variations were observed during other periastron passages as well, the 1997 offers the most complete set, the one that most accurately covers one of two $\Delta DM$ peaks associated with pulsar's passage through the Be star disc. 

$\Delta DM$ was derived from observations obtained at frequencies 1.2, 1.4, 1.5, 4.8 and 8.4~GHz using the Parkes 64-m radio telescope \citep{johnston2001}. The duration of a typical observation was 1 h. The signal was sampled with an interval 0.6 ms and pulse profiles were produced with up to 1024 bins per period. $\Delta DM$ was calculated from the relative difference in the timing residuals between different frequencies. Errors are small $\sim 0.1 \textrm{~pc~cm}^{-3}$. Dispersion measure variations from 13 years of timing observations including 1997 periastron passage are presented in \citet{wang2004}. In 1997 the last detection before the pulsar's eclipse happened 18~days before the periastron passage. There are no multi-frequency $\Delta DM$ data up to $\sim$ 20th~day prior to the periastron passage. The $\Delta DM$ measurement for the orbital phase closer to periastron was obtained for the 16th~day after the 2004 periastron passage, although on this date PSR~B1259$-$63 was not detected at 1.4~GHz \citep{johnston2005}. The value of that measurement is $\Delta DM = 3.2 \textrm{~pc~cm}^{-3}$. 

\begin{figure}[t]
   \centering
   \includegraphics[width=\hsize]{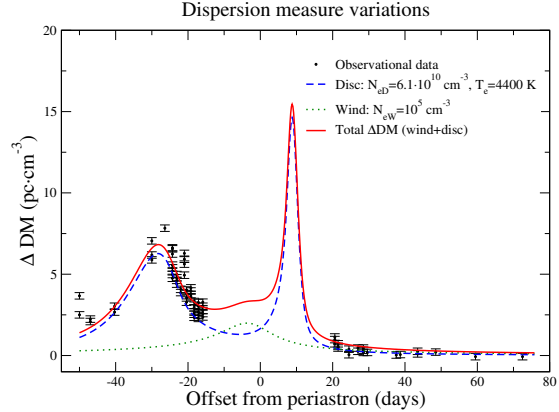}
   \caption{Changes of $DM$ caused by pulsar's orbital motion (points and solid lines correspond to observational values for the 1997 periastron passage \citep{wang2004} and simulated data respectively). The presented simulation results include total $\Delta DM$, as well as the individual $DM$ contribution from the stellar wind and equatorial disc. Observational offset is relative to the value of $146.6 \textrm{~pc~cm}^{-3}$, while the simulated offset was calculated for the minimal value of $DM$, here corresponding to 564th~day after the periastron passage.}
    \label{FigDeltaDM}
    \end{figure}

Originally, the observed $\Delta DM$ values (see the data points in Fig.~\ref{FigDeltaDM}) were measured relative to the value of $146.8 \textrm{~pc~cm}^{-3}$ \citep{connors2002}, which should represent the amount of $DM$ that is generated outside of the binary system. Adopting this external $DM$ value, however, resulted in obtaining slightly negative $\Delta DM$ values. The new mean offset calculated by \citet{wang2004} equals $-0.2 \pm 0.1 \textrm{~pc~cm}^{-3}$, hence, to compare the observational data to our simulations, we increased the observed values by this slight difference.

In our $\Delta DM$ model, we obtained distinctive dual-peak shaped curve related with the high inclination of the disc with respect to orbital plane (see Fig.~\ref{FigDeltaDM}). The disc thickness increases exponentially with the increasing distance from the Be star (see Eq.~\ref{Ne_disc}) and, apparently, when the pulsar approaches this extended part of the disc, the amount of additional $DM$ along a given line of sight increases significantly. As the pulsar moves closer to the star, it also moves towards the thinner part of the disc\footnote{disc thickness decreases exponentially towards the star, while the disc density in the plane of the disc rises ``only'' as $R^{-m}$} the total amount of free electrons along the line of sight actually decreases, at least until the pulsar crosses the disc itself. The other  $\Delta DM$ peak that is shown by the simulation, appears when the pulsar emerges from below the disc, several days after the periastron passage, although we have to mention that there is no observational data to confirm its existence.

The model was fitted to the observational dispersion measure variations by applying the method of the chi-square function minimization. The free parameters were combined in pairs, for each pair the rest of parameters were fixed. We set as the initial values the results of our fine-tuning of the model parameters to the observational data and found the quasi-global $\chi ^2$ minimum for every pair. Next we treated the estimated results as the initial values and performed the chi-square function minimization again, repeating this action several times. The error values are set to $1 \sigma$. Figure~\ref{FigDeltaDM} shows the fitting results in comparison to the observational data. The best fit was obtained for the inclination of the disc (with respect to the orbital plane) of $i_{D} = 61.1^{\circ}\pm ^{3.6}_{3.3}$ and for the disc longitude angle $\omega = 64.5^{\circ}\pm^{3.0}_{4.1}$ (here $\omega$ is the angle between the direction of periastron and the intersection of the Be disc with the orbital plane). The geometry of the system (including the disc density profile parameter $m$)  is mostly responsible for the shape and the lack of symmetry in the $\Delta DM$ peaks, while the amount of the extra $DM$ (i.e. the $DM$ variation scale) allowed us to estimate the characteristic values of the electron density for both the disc and the wind component.

The comparison between the shape and height of the observational and simulated $\Delta DM$ peaks enabled to set initial values of the disc electron density $N_{e D}(0)$, electron temperature $T_{e D}$ (which influences the disc shape by the means of changing the velocity of sound, see Eq.~\ref{Disc_height}) and the disc shape index $m$. These parameters determine Be disc size and shape.
As a result of the fitting procedure, we obtained the following Be disc parameters: $N_{e D}(0) = (6.1 \pm 1.5) \cdot ~10^{10} \textrm{~cm}^{-3}$, $T_{e D} = 4400 \textrm{~K}$, $m=2.86$. 

The disc component of the model has a significantly stronger impact on the simulated $DM$ variations than the stellar wind component, therefore, the scale of the shape of the $DM$ profile is much more dependent on the density of the disc than on the electron density of the wind. This can be seen clearly in Fig.~\ref{FigDeltaDM}, where the contributions to the observed $\Delta DM$ are shown separately for the disc and the wind (as the dashed and dotted curves, respectively). 

The previous, simplified model, which included only the stellar wind \citep{dembska2015} indicated that the thermal absorption in the LS~2883 wind can explain the PSR~B1259$-$63 spectra evolution, provided that the electron density is approximately $N_{e W}(1 \textrm{AU}) = 3.2 \cdot 10^{6} \textrm{~cm}^{-3}$ for the wind temperature $T_{e W} = 10^4 \textrm{~K}$. That density was much higher than the limit $\approx 10^5 \textrm{~cm}^{-3}$ resulting from the current $\Delta DM$ simulations. This is because the wind's $DM$ contribution dominates the disc only for orbital phases very far from the periastron (roughly outside the $-$50 to $+$30~days range), and the $DM$ deviations observed in that range are relatively small, i.e., no larger than $0.5 \textrm{~pc~cm}^{-3}$.

%-------------------------------------------------------------------

\subsection{Initial simulations of PSR~B1259$-$63 spectra thermal absorption}

In the next step, we set the stellar wind electron density equal to estimated from $\Delta DM$ simulations upper limit of $N_{e W}(1 \textrm{AU}) = 10^5 \textrm{~cm}^{-3} $ and we made an attempt to reconstruct the observed high-frequency turnovers for various pulsar orbital phases. The initial simulations were consistent with the observational data for the orbital phases, where the influence of the Be star disc on the pulsar's radiation was dominant, i.e. roughly from $-$30 to $+$30~days around the periastron passage. However, the simulated radio spectra obtained for the given wind density and, for the days outside of the mentioned range, closely resembled a typical power-law spectrum, and no sign of any absorption was visible. This was in contradiction to the observational data that were presented by \cite{dembska2015}.

   \begin{figure}[!htb]
   \centering
   \includegraphics[width=\hsize]{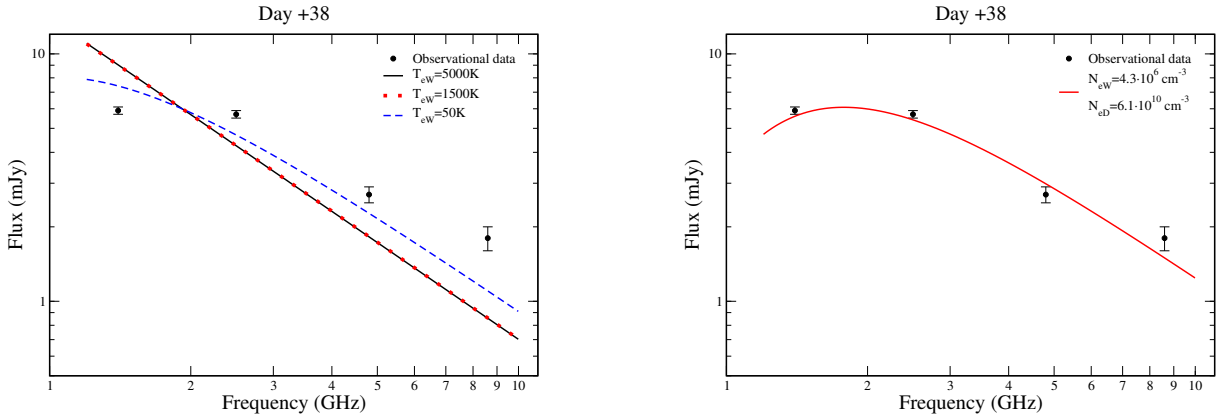}
   \caption{Simulated B1259$-$63 spectra for the 38th~day after the periastron passage (solid lines). The observational data are marked with points. Applied value of wind electron density $N_{e W}(1 \textrm{AU})=10^5  \textrm{~cm}^{-3}$ is the upper limit for $N_{e W}(1 \textrm{AU})$ resulting from modelling of the $DM$ variations. Results presented in the plot correspond to the wind temperatures: $T_{e W} = 5000 \textrm{~K}$, $T_{e W} = 1500 \textrm{~K}$, $T_{e W} = 50 \textrm{~K}$. Even the lowest temperatures do not cause significant absorption in the stellar wind for the chosen electron density.}
    \label{FigWrongFitGoodDM}
    \end{figure}
    
   \begin{figure}[!htb]
   \centering
   \begin{minipage}{\hsize}
   \includegraphics[width=\hsize]{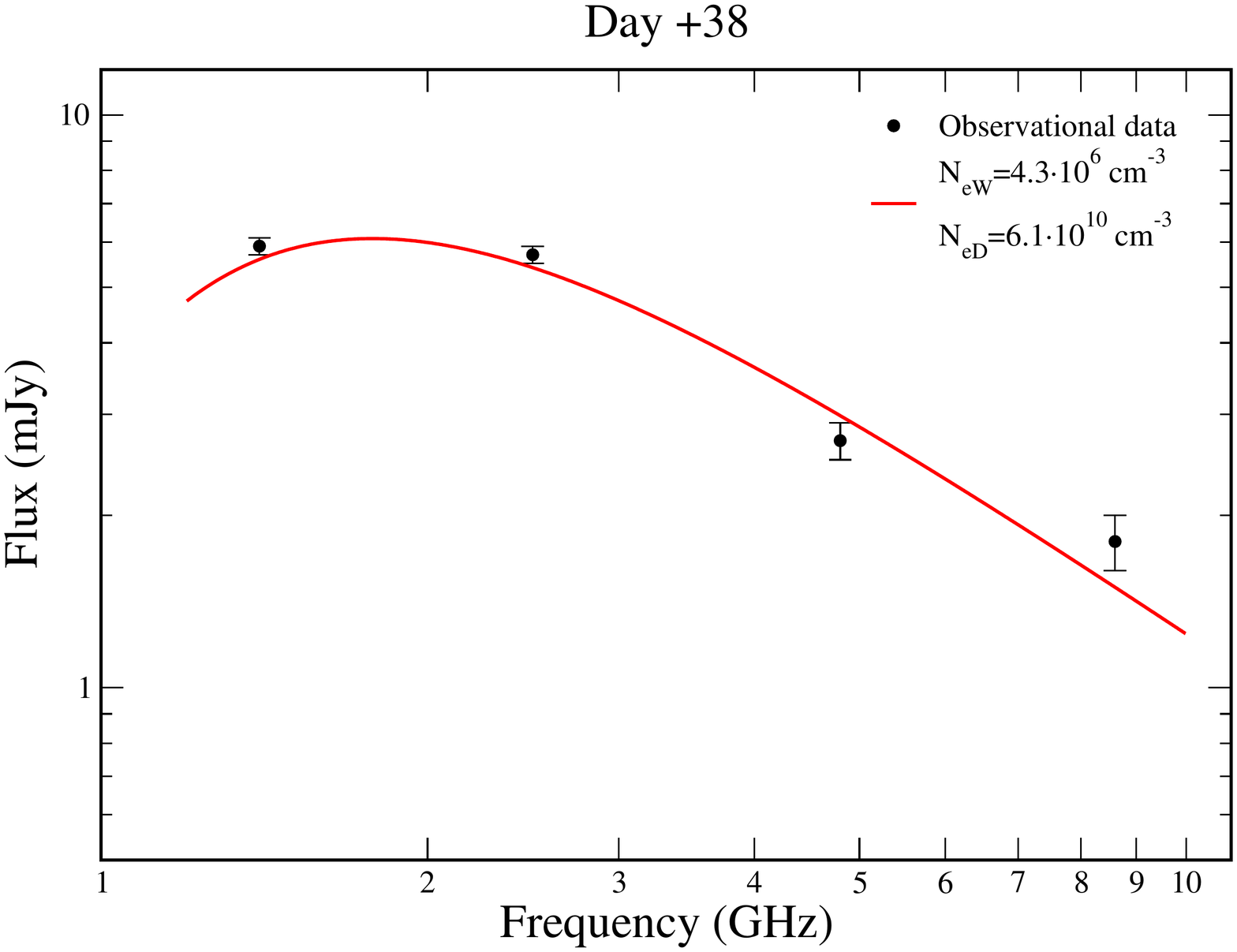}
   \vspace{0.3cm}\\
   \includegraphics[width=\hsize]{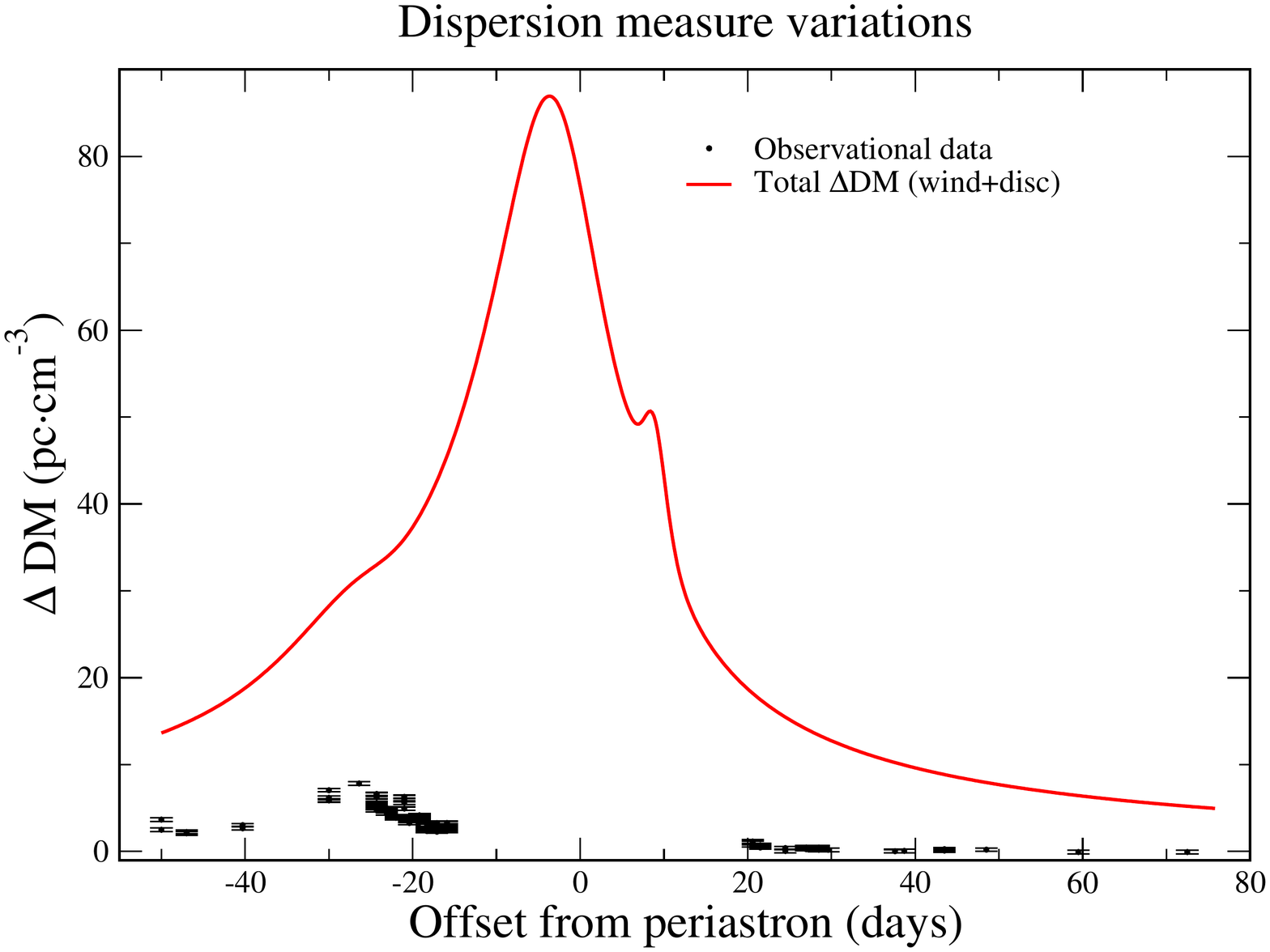}   
   \end{minipage}
   \caption{In the upper panel we present the result of PSR~B1259$-$63 spectrum absorption simulation for the 38th~day after the periastron passage. Here we neglect the value of the wind electron density $N_{e W}(1 \textrm{AU})=10^5  \textrm{~cm}^{-3}$, which we concluded from $\Delta DM$ computations. Applying $N_{e W}(1 \textrm{AU})=4.3\cdot10^6  \textrm{~cm}^{-3}$, we get a  satisfactory level of the spectrum absorption, at the same time losing agreement with $\Delta DM$ observations, as could be expected (lower panel).}
    \label{FigGoodFitWrongDM}
    \end{figure}       

Fig.~\ref{FigWrongFitGoodDM} shows the result of our simulated spectrum for the 38th~day after the periastron passage, compared to the observational data from that particular day. The observed spectrum (data points), does not exactly follow a power-law, the flux at the lowest frequency seems to be too low, indicating a turnover, or at least a break in the spectrum. Simulated data, however, (dotted line) are represented by a straight line, suggesting a power-law spectrum with no visible absorption signatures. 

For the orbital phase corresponding to this observing epoch (and other epochs outside the $-$40 to $+$30 day range), there is no adjustment to the stellar disc parameters in our model, either physical or geometrical, that would provide enough absorption to explain the observed shape of the spectrum without simultaneously destroying the $DM$ variations agreement (see Fig.~\ref{FigDeltaDM}). Therefore, the only logical way would be to adjust the parameters of the stellar wind.

The amount of the observed absorption depends on both the electron density and the temperature. The absorption gets stronger for lower electron temperatures, but changing the temperature of the absorber does not affect the amount of $DM$ it contributes. However, as it is indicated by the dashed curve in Fig.~\ref{FigWrongFitGoodDM}, even decreasing the temperature down to 50~K does not fully satisfy the fit to the observational data. Obviously, since we postulated that the absorption takes place in the outflowing stellar wind of a Be star, such low values of the electron temperature are un-physical and not acceptable.

The other possibility to increase the amount of the observed absorption is to increase the electron density. The top plot of Fig.~\ref{FigGoodFitWrongDM} shows a simulated spectrum obtained for the disc wind temperature of 10000~K (which should be a typical electron temperature for a Be-star wind), but for the electron density $N_{e W}(1 \textrm{AU})=4.3\cdot10^6 \textrm{~cm}^{-3}$, i.e. about 50 times higher than our analysis of the $DM$ variations indicated. This spectrum roughly agrees with the observed pulsar flux densities, however, using this value to recreate the $DM$ values makes the simulated $DM$ variations to be to large, by almost an order of magnitude, which is shown in the bottom plot of Fig.~\ref{FigGoodFitWrongDM}.

%-------------------------------------------------------------------

\subsection{A modified model}

Since neither the influence of the stellar disc (which does not affect the pulsar radiation when it is far from the periastron passage), nor the stellar wind can explain the observed turnovers and the clearly non-power-law shapes of the pulsar spectra observed far from the periastron passage (at least without affecting the simulated $\Delta DM$ values), there is only one possibility: an external absorber. Such an absorber, in principle, has to be located outside the pulsar's orbit, so it will contribute equally to the spectra for all orbital phases. As for the $DM$ contribution, it will provide only a constant value, and hence it will not influence the $DM$ variations (since the total ``external'' $DM$ is subtracted from the observed values anyway).

Using such a hypothesis, we were able to estimate the parameters of the external absorber based on the spectra of PSR~B1259$-$63 for orbital phases far from the periastron passage. The spectral analysis, however, only gives the total amount of absorption, not the physical parameters.

For solitary pulsars, the role of such external (and the only) source of absorption, may be fulfilled by different kinds of objects. \cite{lewandowski2015}, proposed that the absorber might be an extensive H~II region, a pulsar wind nebula around the neutron star, or a dense supernova remnant filament. For isolated GPS pulsars these ideas were explored by \cite{kijak2017}. 

Constraining the physical parameters of the absorber is a difficult task without additional observational information, such as the size of the absorber, its temperature or density. Most of the time one can only use the total $DM$ observed for the pulsar to put some boundaries on the absorber's density, since the amount of $DM$ contributed by an absorber cannot exceed the total value of $DM$.

In Table~\ref{TabExtraAbsorber}, we present a few of the possible configurations of the absorber's parameters, assuming different electron temperatures, densities and sizes - all contributing $100 \textrm{~pc~cm}^{-3}$ to $DM=146.6 \textrm{~pc~cm}^{-3}$ which is the total ``external'' dispersion that provides a hard limit for our calculations. All these sets of parameters will produce the same amount of absorption, the values of the optical depth given in the table are calculated for the observing frequency of 1.4~GHz.

These parameters were obtained by manual fitting the simulated spectra to the observed flux measurements. We have chosen not to use any formal mathematical algorithm, since we had to satisfy almost 20 observational data sets (observed spectra for orbital phases prior to day $-40$ and past $+30$) with the same number of simulated spectra, and in addition to the absorption-related parameters, the intrinsic pulsar spectrum scaling parameter ($A$ in Eq.~\ref{power-law}) also had to be a free parameter of the fit for each of the observed spectra\footnote{the spectra of the pulsar are apparently heavily affected by interstellar scintillation, as it was shown by \citet{kijak2011a}.}.  That would result with a modelling algorithm with almost 30 free parameters, and most probably prove futile. Instead, the parameters of the absorber were adjusted so as the shapes of the simulated spectra resembled the observational data as close as possible, and then the flux scaling parameter $A$ was adjusted for every individual spectrum.

\begin{table}[!htb]
\small

\begin{tabular}{p{0.2\linewidth}p{0.2\linewidth}p{0.2\linewidth}p{0.2\linewidth} }
\tableline
EM&$T$ & $N_{e }$ & $R$\\
$(\textrm{pc} \cdot \textrm{cm}^{-6})$ &  $ (\textrm{K})$ & $ (\textrm{cm}^{-3})$ & $\textrm{(pc)}$\\
\tableline
\rule{0pt}{3ex}
$10^{7}$ &15200 & $10^{5}$ & 0.001\\
~$10^{6}$ & 2770 & $10^{4}$ & 0.01\\
~$2\cdot 10^{5}$ & 820 & $2 \cdot 10^{3}$ & 0.05\\
~$10^{5}$ & 477 & $10^{3}$ & 0.1\\
~$10^{4}$ &70 & $10^{2}$ & 1\\
\tableline
\end{tabular}

\caption{Example sets of derived parameters for the additional absorber, assuming the optical depth of $\tau_{\textrm{1.4GHz}} = 0.97  \textrm{~pc} \cdot \textrm{cm}^{-6}$ and the $DM$ contribution from the absorber of $100$~pc~cm$^{-3}$.}
\label{TabExtraAbsorber}
\end{table}

\begin{table}[!htb]
\vspace{0.5cm}
\huge
\resizebox{\columnwidth}{!}{%
\begin{tabular}{lcr}
&B1259$-$63/LS~2883&\\
\tableline
parameter name&parameter
 value&references\\
\tableline
&B1259$-$63&\\
& orbital parameters&\\
\tableline
\rule{0pt}{3ex}
period $P_{b}$&1237~days&(1)\\
~inclination $i$&36$^{\circ}$&(1)\\
~eccentricity $e$&0.87&(1)\\
~longitude of periastron $\omega$&138.65$^{\circ}$&(1)\\
~semi-major axis $a$&4.42 AU&$a \sin{i} = 2.60 \textrm{AU}$ (1)\\
\tableline
&Be wind parameters&\\
\tableline
\rule{0pt}{3ex}
$T_{e W}^\star $&10000~K&10000~K (2), (10)\\
~$N_{e W}(1 \textrm{AU})^\star $&$10^{5} \textrm{~cm}^{-3}$&$^*$This paper\\
&&\\
\multicolumn{3}{p{3\linewidth}}{~~~~~~~~~~~~~~~~~~~~~~~~~~$N_{eW}(r)$ derived from $N_{eW}(r) \propto r^-2$}\\
&&\\
\rule{0pt}{3ex}
$N_{e W}(50~R_*)$&$ 5.2 \cdot 10^{5} \textrm{~cm}^{-3}$& 
$7.3\cdot 10^5  \textrm{~cm}^{-3} (3)$\\
~$N_{e W}(1~R_*)$&$1.3 \cdot 10^{8} \textrm{~cm}^{-3}$&
$ 2\cdot10^9  \textrm{~cm}^{-3}$ (4)\\
\tableline
&Be disc parameters&\\
\tableline
\rule{0pt}{3ex}
inclination $i_{D}^\star $&$61.1^{\circ}\pm ^{3.6}_{3.3}$&$10^{\circ}-40^{\circ}$ (4)\\
&& $>30^{\circ}$ (5)\\
&& $\sim 35^{\circ} \pm 7 $ (6)\\
~longitude angle $\omega_{D}^\star $&$64.5^{\circ} \pm ^{3.0}_{4.1}$&$\sim 90^{\circ}$ (7)\\
&&$\sim 70^\circ (8)$\\
~shape parameter $m^\star $&$2.86 \pm ^{0.6} _{0.4}$&2-3.5 (9)\\
\rule{0pt}{3ex}
$T_{e D}^\star $&$4400 \textrm{~K} \pm ^{3100}_{2100} $&4000~K (10)\\
&&$13680 \textrm{~K }$ (11)\\
\rule{0pt}{3ex}
$N_{e D}(0)^\star $&~$(6.1 \pm 1.5) \cdot 10^{10} \textrm{~cm}^{-3}$~&~$10^8 - 10^{10}  \textrm{~cm}^{-3} $ (4)\\
&&$\sim 10^{11} \textrm{~cm}^{-3} $ (8)\\
\tableline
&Additional absorber&\\
\tableline
\rule{0pt}{3ex}
size $d^\star$&0.01~pc&$^*$This paper\\
~$N_{e}^\star$&$10^{4} \textrm{~cm}^{-3}$&$^*$This paper\\
~$T_{e}^\star$&2770~K&$^*$This paper\\
\tableline
\end{tabular}
}
\caption{Physical and geometric parameters used in B1259$-$63/LS~2883 model. Free parameters are denoted with symbol $^\star$. {Derived $N_{eW}(r)$ parameters resulted from the adopted free parameter value $N_{eW}(1 \textrm{ AU})$.} References: (1) - \citep{johnston94}, (2) - \citep{waters92}, (3) - \citep{johnston2001}, (4) - \citep{melatos95}, (5) - \citep{wex98}, (6) - \citep{shannon2014}, (7) - \citep{johnston99}, (8) - \citep{chernyakova2006}, (9) - \citep{waters86}, (10) - \citep{dachs87}, (11) - \citep{okazaki2011}.}
\label{TabAllParams}
\end{table} 
%%Boe.

For the purposes of the following analysis, we used an external absorber with the size of 0.01~pc, the electron density $N_{e}=10^4$~cm$^{-3}$, and temperature $T_{e} = 2770$~K (although it should be highlighted, that we would obtain the same results by adopting any parameter set presented in Table~\ref{TabExtraAbsorber}). Such an absorber provides 100~pc~cm$^{-3}$ towards the total $DM$ value, its emission measure is $EM=10^6$~pc~cm$^{-6}$ and this results in the optical thickness of $\tau_{1.4 \textrm{GHz}} \approx 0.97$. For larger absorbers the electron density has to decrease to keep the total DM contribution at the assumed value. However, since the emission measure is proportional (in the case of a uniformly dense absorber) to $N_{e}^2$, this causes a huge drop of the optical depth, that can be compensated only by lowering of the electron temperature. For larger absorbers of the size of about 1 parsec the temperature needed to explain the observed absorption would have to be as low as 70~K. This value is clearly to low to the degree of being unphysical for an ionized region of the ISM. 

\subsection{Final model results}

\begin{figure}[!t]
   \centering
   \begin{minipage}{\hsize}
      \centering
   \includegraphics[width=\hsize]{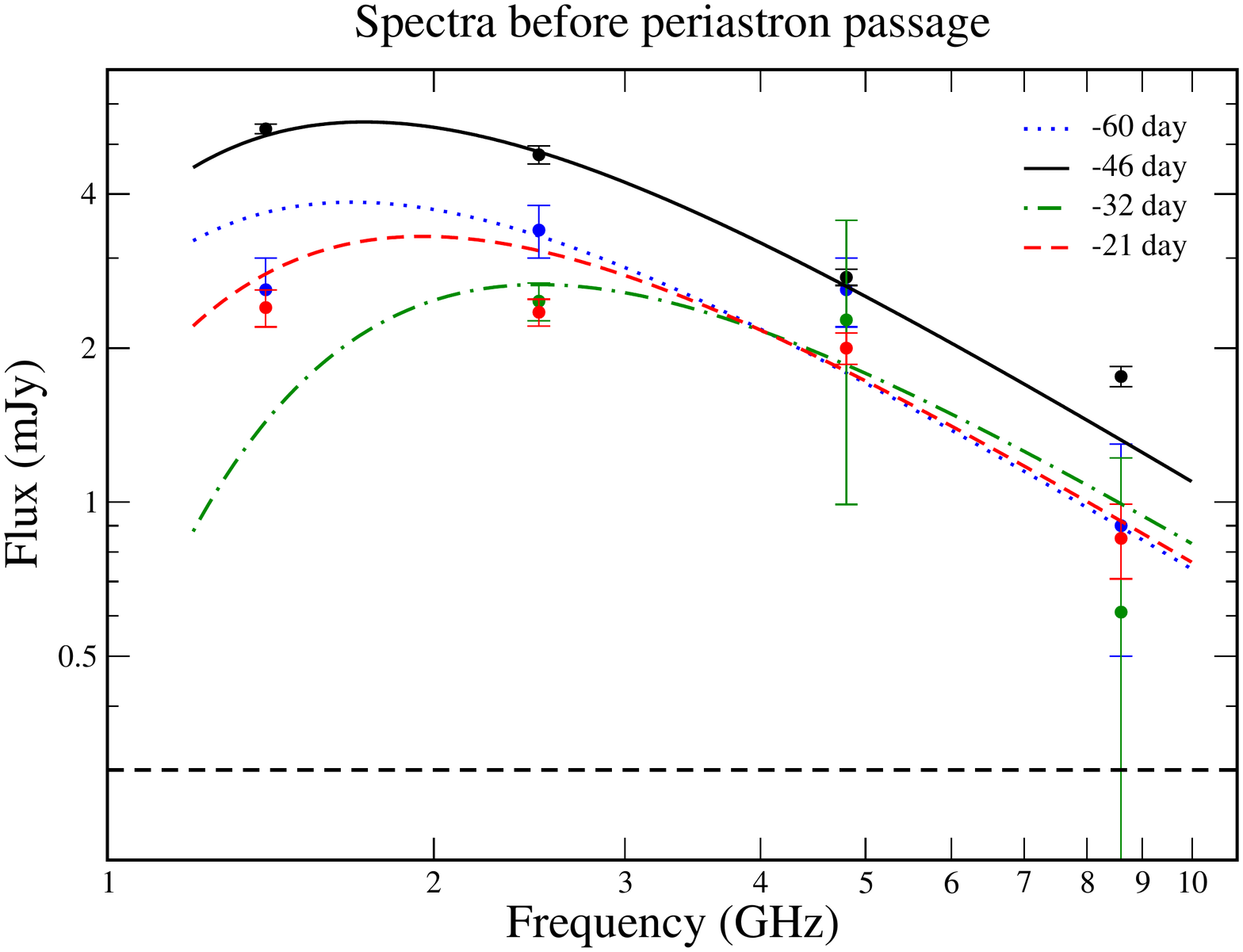}
   \includegraphics[width=\hsize]{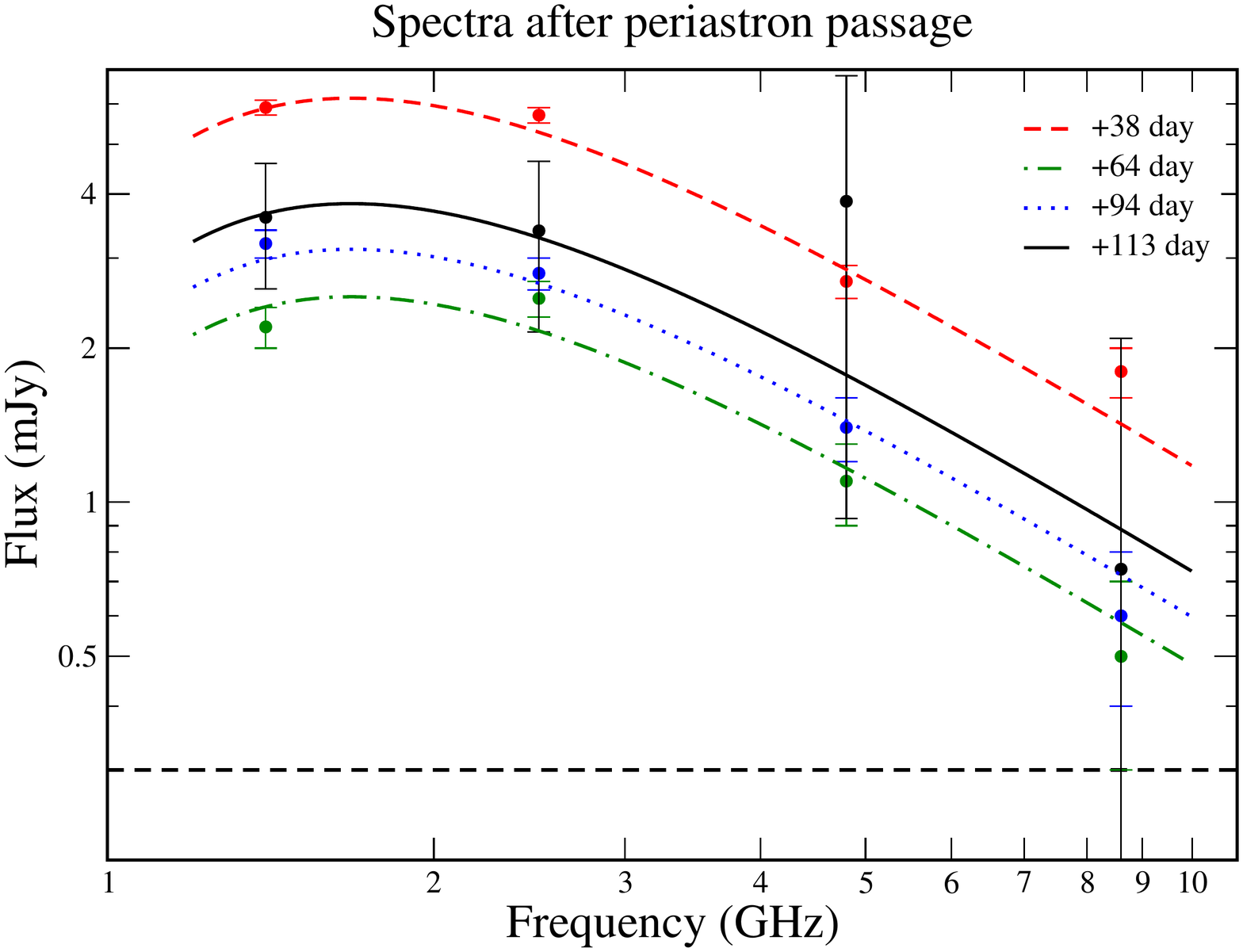}
   \end{minipage}
   \caption{Spectra resulted from modelling the influence of the free-free absorption in the LS~2883 stellar wind and equatorial disc for the PSR~B1259$-$63 radio emission. The upper panel: Strong absorption for the $-32$~day (dash-dotted line) is caused by the stellar disc, which reaches the maximum value in the first peak of $\Delta DM$ for the day $=-29$. The lower panel: Here the disc influence on the pulsar radiation is not relevant. The observational spectra show weak absorption that cannot be caused by the stellar wind with electron density $N_{e W}(1 \textrm{AU})=10^5  \textrm{~cm}^{-3}$. The spectra shapes presented here were modeled by taking into account an additional absorber with the electron density $N_{e}=10^4 \textrm{~cm}^{-3}$, temperature $T = 2770 \textrm{~K}$ and size $d = 0.01 \textrm{~AU}$. The horizontal line $S_\nu = 0.3 \textrm{ mJy}$ represents the detection limit from \citet{johnston99}. } 
    \label{FigFinalResults}
    \end{figure}
    
After the inclusion of the external absorber in our model, we proceeded with the modelling of the spectra observed close to the periastron passage, where the observed amount of the thermal free-free absorption is affected by all of the model components: the circumstellar disc, the stellar wind and the external absorber. The geometrical parameters of the disc, as well as the characteristic wind and disc electron densities, were set to the values obtained from the $DM$ variations model. In principle, the disc temperature was also estimated from that fit, since it affects the shape of the disc (see Eq.~\ref{Disc_height}), however, the resulting constrains on the temperature are quite weak. %On the other hand, the same temperature is also heavily affecting the amount of absorption provided by the disc, hence, we decided to treat it as a free parameter when modelling the pulsar spectra. The electron temperature of the stellar wind was also a free parameter of the fit.%

In the modelling of the spectra, we treated each observing day epoch separately, namely, for each spectrum we fitted a separate intrinsic flux scaling parameter $A$ (in Eq.\ref{power-law}). It can be easily noticed in the observational spectra (see \citealt{kijak2011a} and \citealt{dembska2015}) that the overall flux density level of the spectra changes considerably from observation to observation. This is probably due to the effects of interstellar scintillations, most likely the refractive scintillations which at higher frequencies are usually characterized with a relatively small time scale of the order of weeks or days\footnote{see for example the observations of PSR~B0329+54 by \cite{lewandowski2011}, where the refractive time scale at the frequency of 5~GHz is close to 5 hours. While the $DM$ of that pulsar is significantly lower (26 versus 146.6~pc~cm$^{-3}$) and the distance to B1259$-$63 is slightly more than twice as high (1.0~kpc versus 2.3~kpc), the dependence between the distance and the refractive time scale is relatively moderate, as $t_{RISS}\propto d^{1.6}$. This could mean that the refractive time scale for B1259$-$63 is of the order of a few days at most at 5~GHz, and even lower at 8~GHz}. In our model, we had to compensate for that effect using different flux scaling factors for every spectrum. As an effect, for each of the simulated spectra, their shapes were determined by the physical and geometrical parameters of the model components, while their scale was adjusted independently for each spectrum.

Figure \ref{FigFinalResults} shows the simulated radio spectra for selected observational epochs corresponding to the orbital phases prior to the periastron passage (the upper plot) and past it (lower plot). Looking at the plots, one can still find a few outliers (which, again, are probably caused by scintillation effects), but our model attempts to fit all the spectral shapes simultaneously, hence the shapes of the individual spectra cannot be adjusted separately to conform the observational data precisely. Overall, we found our simulated data with a relatively good agreement to observations.
Table \ref{TabAllParams} summarizes all physical and geometric parameters obtained using our model that satisfy both the $DM$ variations observed during the 1997 periastron passage, and the spectra observed for all the observing epochs. In the last column of the Table \ref{TabAllParams}, we compare the derived values of the free parameters with data found in literature. Parameter $T_{e W}$ has no impact on the simulations results, since the $\Delta DM$ doesn't depend on the wind temperature and the stellar wind of density $N_{e W}(1 AU) \le 10^5 \textrm{~cm}^{-3}$ is to weak to have effect on the decrease in the pulsar spectra. Therefore, we couldn't estimate its electron temperature $T_{e W}$, and we set its value to $T_{e W} = 10000 \textrm{~K}$ \citep{waters92}. The disc temperature $T_{e D} = 4400 \textrm{~K} \pm ^{3100}_{2100} $ resulting from our modelling is much lower than $0.6~ T_{eff} = 13680$~K assumed by \citet{okazaki2011} for the isothermal, optically-thin disc model. However, the Be star discs are strongly non-isothermal with a significantly cooler inner optically-thick region \citep{carciofi2006}, with a temperature dropping to several thousand K in the disc's equatorial plane \citep{millar99}. This cool and dense part of the Be star disc affects pulsar spectra and $DM$ variations stronger than its outer layers. The inclination of the Be star disc $i_D = 61.1^{\circ}\pm ^{3.6}_{3.3}$ obtained using our model goes beyond the range $10^{\circ}-40^{\circ}$ estimated by \citet{melatos95}, which was based on a very limited observational sample containing only 7 $\Delta DM$ measurements around the periastron \citep{johnston96}. The advance of the periastron calculated from timing observations of PSR~B1259$-$63 allowed \citet{wex98} to conclude, that the Be star is tilted by more than $30^\circ$ to the orbital plane (e.g. $\sim 75^\circ$ if LS 2883 rotates at 70 per cent of its break-up velocity). The large inclination of the Be star with respect to the orbital plane can be explained by a significant birth kick to the pulsar (e.g. \citealt{wex98}, \citealt{hughes99}). The disc's tilt obtained by \citet{shannon2014} on the base of timing measurements is somewhat smaller $\sim 35^{\circ} \pm 7 $, but this value was calculated taking into account a larger Be star mass $M_* \approx 30 M_\odot$ and a different orbital inclination $i = 23^\circ$.

\begin{figure}[!htb]
   \centering
   \begin{minipage}{\hsize}
      \centering
   \includegraphics[width=\hsize]{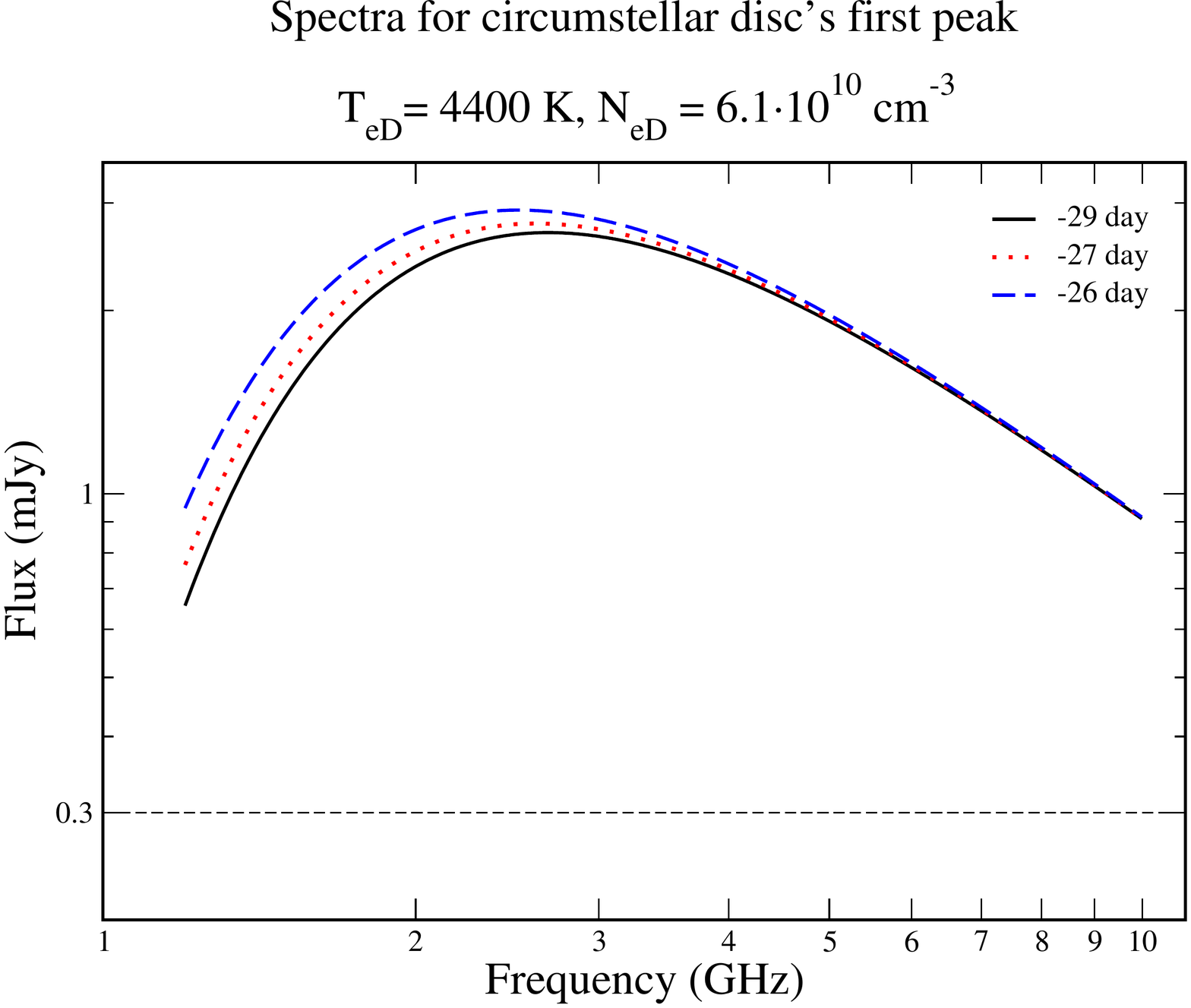}
   \includegraphics[width=\hsize]{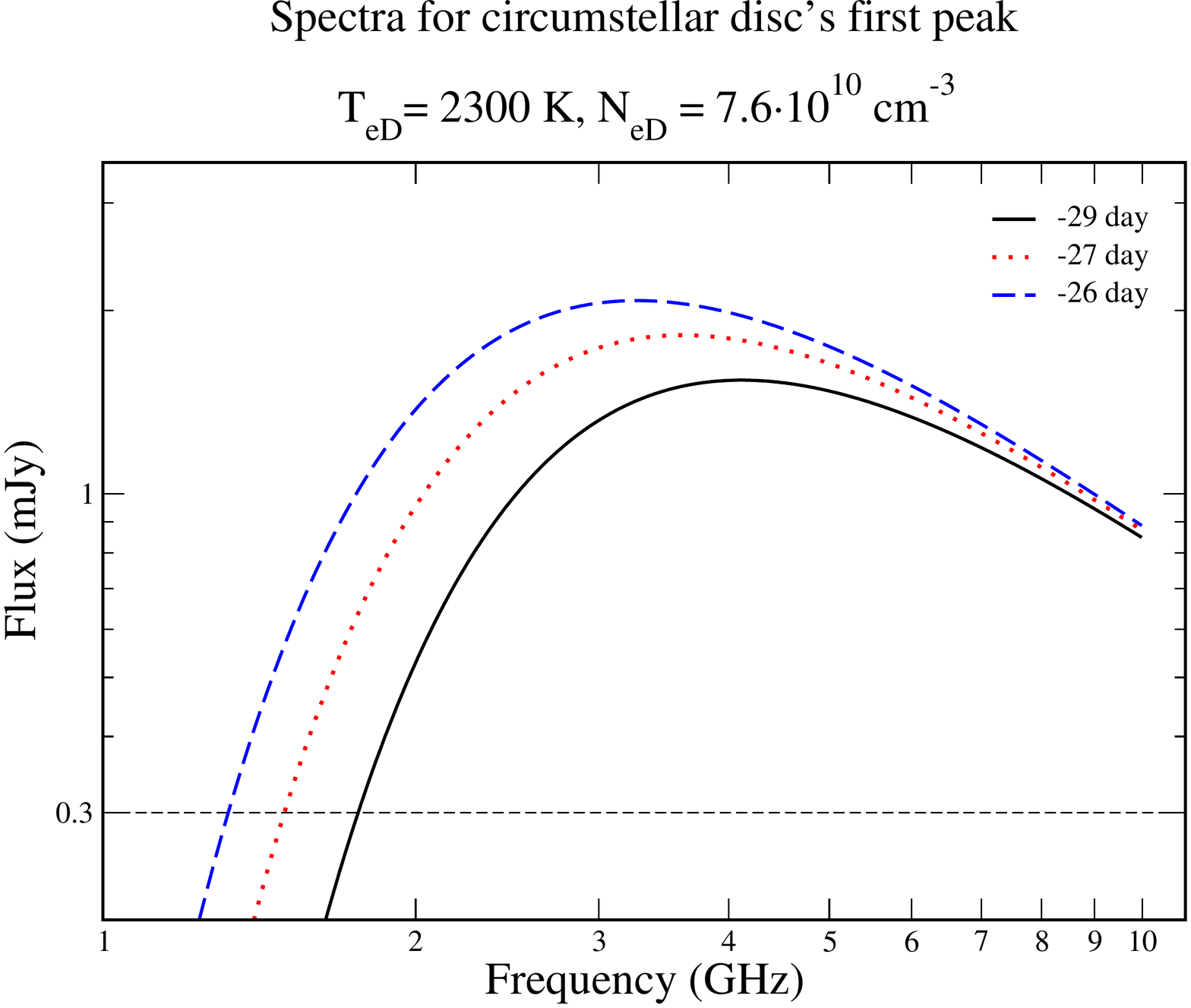}
   \end{minipage} 
   \caption{Simulated spectra for the days around $-$29~day, where the circumstellar disc reaches its maximum value of $DM$ and the thermal absorption before the periastron passage is the strongest. Upper panel: spectra for the best-fit parameters $T_{eD}$ and $N_{eD}(0)$. Lower panel: spectra for the limit values of $T_{eD}$ and $N_{eD}(0)$ in the range of error $=1~\sigma$. For the disc temperatures lower than $\sim$ 3500~K and electron densities greater than $6.7\cdot 10^{10} \textrm{~cm}^{-3}$ the spectrum flux density on the low-frequency side falls rapidly even below the detection limit. The peak frequency is slightly shifting towards lower observing frequencies with the decrease of the disc $DM$. }
    \label{FigDetectionLimit}
    \end{figure}

One of the examples of the explanatory power of our model is the analysis of some of the individual flux measurements, especially at the lowest observing frequency of 1.4~GHz. Looking at the available data, especially during the 1997 periastron passage, there are no L-band flux density measurements for the observing epochs close to $-30$~day, while they are available for the earlier and later epochs (even some of the epochs between the $-30$~day mark and an actual eclipse which happened around day $-20$). This lack of measurements, which we can only explain by non-detections, coincides with the first peak of the $DM$ variations plot (see Fig.~\ref{FigDeltaDM}). The increase of the $DM$ is obviously caused by the increased electron column density, therefore, it should correspond to an increase of the predicted amount of absorption. Indeed, when using our model (set to its final parameters), the simulated spectra clearly show that the expected pulsar flux drops significantly for these epochs (see Fig.~\ref{FigDetectionLimit}) and for the electron temperatures lower than 3500~K (this would be still within $1 \sigma$ level from our estimated best-fit $T_{eD}$ value) the flux falls below the level of 0.3~mJy,  which was quoted by \cite{johnston99} as their detection limit for all the observing frequencies. And, since the pulsar was below the detection limit, it is understandable that there are no flux density measurements for these epochs.  

\subsection{Possible explanations for
 an external absorber}

The binary system B1259$-$63/LS~2883 is likely a member of the Cen OB1 association \citep{negueruela2011}. This association, centered at galactic coordinates $l=303.7^{\circ}$, $b=0.5^{\circ}$, has a relatively large angular size of roughly $4\times 4$ degrees. For this reason, one of the possible candidates for an external absorption will be an HII region, related to the LS~2883 star itself, or to the association. The possible sets for parameters that would satisfy both the observed amount of absorption and dispersion measure, however, indicate that the absorber is rather small in size (see Table~ \ref{TabExtraAbsorber}), much smaller than a H~II region with a typical size of $\sim$ 10~pc and temperatures around 5000$-$7000~K \citep{afflerbach96}. Even when assuming very low electron temperature of 70~K, which is much too low for an H~II region, the size of the absorber cannot exceed 1~parsec. The likely candidate for an external absorber could be a young, ultracompact HII region. This kind of object is characterized by an extremely small size and a high electron density (according to \citealt{franco2000} the ultracompact HII regions have sizes of the order of 0.001~pc and can be even denser than $10^6 \textrm{~cm}^{-3}$. Slightly less extreme values were given by \citealt{wood89}: $N_e > 10^4 \textrm{~cm}^{-3}$, size $<$0.1~pc and temperatures around 1000$-$10000~K.)

Another possibility explored by \cite{lewandowski2015} in their search for explanation of gigahertz-peaked spectra of solitary pulsars, was the absorption in dense but ionized supernova remnant filaments. OB associations usually host a number of supernova remnants, and the same holds for Cen~OB1. There are documented observations of a faint, incomplete shell, with possible extension towards southwest with angular size $\sim 37\times 23$ arcmin, at $l= 301.4^{\circ}$ and $b=  -1.0^{\circ}$ \citep{whiteoak96}. Another, distorted supernova remnant shell, 17 arcminutes wide, is located at $l= 302.3^{\circ}$ and $b=  +0.7^{\circ}$. While neither of these supernova remnants coincides with the PSR~B1259$-$63/LS~2883 system\footnote{$l= 304.18^{\circ}$ and $b=  -0.99^{\circ}$}, we cannot exclude a possibility of another, undetected remnant inside the Cen~OB1 association. Filaments of such an old, relatively cool but photo-ionized remnant, which are typically about 0.1 parsec wide, can provide the necessary amount of absorption to explain the apparent spectra of the pulsar. \cite{lewandowski2015} pointed out that dense filaments in supernova remnants can be the source of enough absorption to cause a high-frequency turnover in a pulsar spectrum. Such a filament would need to have properties similar to the filaments found in the supernova remnant G11.2$-$03 \citep{koo2007}, i.e., the electron density of $\sim 6600 \pm 900 \textrm{~cm}^{-3}$, the temperature of $\sim$ 5000~K and the size of $\sim$ 0.24~pc.  The only difference with our case would be that the filament causing absorption for B1259$-$63 would have to be an order of magnitude thinner and a few times denser than the filaments measured in G11.2$-$0.3.

The last scenario proposed for the isolated GPS pulsars was a cometary shell of a bow-shock pulsar wind nebula, however, we believe that this possibility does not apply for PSR~B1259$-$63, since the wind nebula around it is rather untypical (which is due to extremely high density of the matter within the system), and highly variable with the pulsar orbital motion  (see for example \citealt{chernyakova2014}).
 
%-------------------------------------------------------------------

\section{Conclusions}
        
 We have presented a successful attempt to model the basic physical properties and geometry of the extended components of the binary system of PSR~B1259$-$63 and LS~2883, namely the stellar (polar) wind ejected by the Be star, as well as the hot circumstellar disc. Both of these components affect the outgoing radio emissions of the pulsar, changing their characteristics, providing extra amount of interstellar dispersion and absorbing a part of the pulsar's flux, most likely through the phenomenon of thermal free-free absorption. We have to mention, however, that in our model we have neglected the possible absorption of B1259$-$63 flux caused by the cyclotron resonance in the magnetic field of the Be star disc.
 
The frequency-dependent absorption of the pulsar flux causes the spectrum of the pulsar to turnover at the frequencies close to and above 1~GHz depending on the pulsar's orbital phase, which allows us to classify this object as a member of the group of gigahertz-peaked spectra (GPS) pulsars.

Contrary to our earlier attempts \citep{dembska2015}, our model suggests that most of the absorption takes place not in the stellar wind of the Be star, but rather in an extended and relatively thick disk around the companion, and the same can be said for the observed dispersion measure variations. To satisfy both the $DM$ variations and the observed flux density spectra, however, we had to introduce an external absorber of the pulsar radiation to our model. This external absorption is probably caused by the environment of the system, which is a member of the Cen~OB1 association, and comes either from a compact and dense H~II region or a supernova remnant filament.

%--------------------------------------------------------------------

\section*{Acknowledgments}
We thank the anonymous referee for the comments and suggestions that helped us to greatly improve the paper.
This work was supported by the grant DEC-2013/09/\linebreak B/ST9/02177 of the Polish National Science Centre.

\end{document}